\documentclass[journal]{IEEEtran}
\usepackage{graphicx}
\usepackage{t1enc}
\usepackage{rotating}
\usepackage{amsmath}

\ifCLASSINFOpdf
\else
\fi

\begin{document}
\title{Performance evaluation of the QOS provisioning ability of IEEE 802.11e WLAN standard for multimedia traffic }

\author {Venkata Sitaram. A,~ 
        Venkatesh. T. G, ~Arun George,
         ~Manivasakan. R, ~and ~ Bhasker Dappuri.% <-this % stops a space
\IEEEcompsocitemizethanks{\IEEEcompsocthanksitem Department
of Electrical Engineering, Indian Institute of Technology, Madras,
India, 600036.\protect\\ }}

\maketitle

\begin{abstract}

This paper presents an analytical model for the average frame transmission delay and the jitter for the different Access Categories (ACs) of the IEEE 802.11e Enhanced Distributed Channel Access (EDCA) mechanism. Following are the salient features of our model. 
 As defined by the standard we consider (1) the virtual collisions among different ACs inside each EDCA station in addition to external collisions.
(2) the effect of priority  parameters, such as minimum and maximum values of Contention Window (CW) sizes, Arbitration Inter Frame Space (AIFS). (3) the role of Transmission Opportunity (TXOP) of different ACs. (4) the finite number of retrials a packet experiences before being dropped. Our model and analytical results provide an in depth understanding of the EDCA mechanism and the effect of Quality of Service (QoS)  parameters in performance of IEEE 802.11e protocol.
\end{abstract}


\IEEEpeerreviewmaketitle

\section{Introduction}
\label{intro}

The IEEE 802.11  standard and many of its enhancements like  a, b, e, g, n, ad are  some of the most widely deployed  Wireless Local Area Network (WLAN) standards \cite{IEEEhowto:kopka}.   These days IEEE 802.11 WLAN based hotspots are available at offices, university campuses, airports, hotels, residential areas etc. enabling people to get connected to the Internet easily.   The WLAN can be set up in a  simple manner in almost all areas without requiring the use of extensive infrastructure. 

 The IEEE 802.11 WLAN  defines two architectures namely Basic Service Set (BSS) and Independent Basic Service Set (IBSS).  In BSS all wireless nodes are connected to an  access point (A.P.) and communicate only through that A.P. In (IBSS) nodes communicate  directly with each other in an adhoc manner i.e without an A.P. 

%\subsection{Overview of IEEE 802.11 MAC Protocol}

IEEE 802.11 MAC defines two different mechanisms for frame transmission namely  Distributed Coordination Function (DCF) and Point Coordination Function (PCF).   DCF uses  Carrier Sense Multiple Access with Collision Avoidance (CSMA/CA) mechanism with slotted Binary Exponential Back-off (BEB) algorithm. PCF provides controlled channel access through polling. 

Today's WLAN carry many forms of media such as video, voice, data and signaling.   Information services on the Internet come in varying forms, such as web browsing, e-mail, VOIP based telephony and multimedia on demand. This means that  different applications have different service features and desired QoS. Multimedia applications such as VOIP, and video-on demand have stringent QOS requirements and are sensitive to delay and jitter in their traffic. In contrast web-browsing are bursty in nature and do not require QOS service guarantee on delay and jitter. While multimedia traffic can tolerate packet loss, applications like FTP needs to be very reliable.

  In the IEEE 802.11 standard, there is no traffic differentiation. In view of this an enhanced Channel access mechanism has been introduced in  the IEEE 802.11e standard to satisfy the QoS requirements of different traffic. The EDCA mechanism has also been implemented in ieee 802.11p vehicular Networks \cite{5712769} to maintain traffic priorities.
  
  In this work we carry out the performance analysis of IEEE 802.11e standard to bring out its ability to provide QoS for the multimedia traffic. To this end we first give an overview of the IEEE 802.11 and 802.11e MAC protocol followed by the literature survey before doing the analysis. 

\subsection{Overview of IEEE 802.11 and 802.11e MAC Protocol}

\begin{figure}
  \centering
  \includegraphics[width=3in,height=1.2in,angle=0]{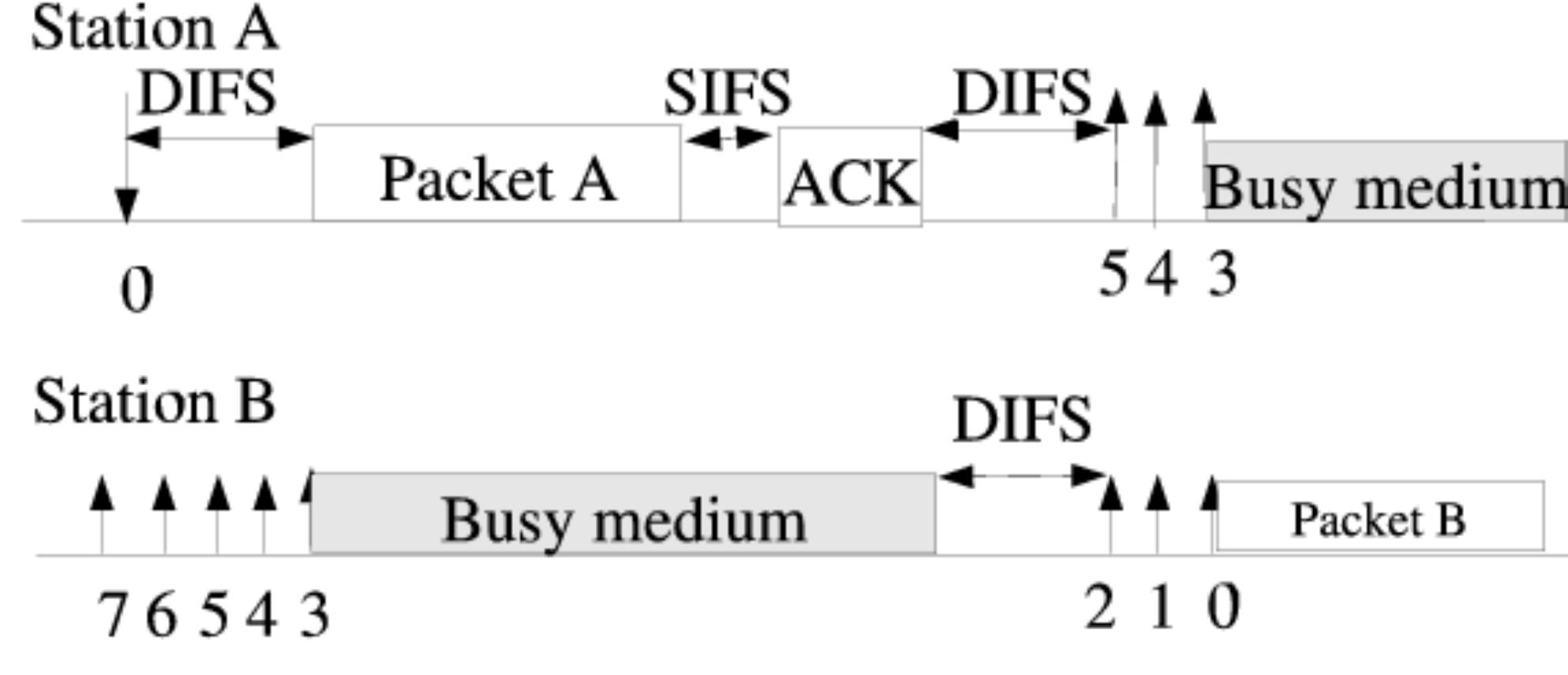}
\caption{Basic Access mechanism}
\label{acsqos1}
\end{figure}%
%\end{center}

\begin{figure}
\includegraphics[width=3.5in,height=2.25in,angle=0, trim=0mm 0mm 0mm 0mm, clip]{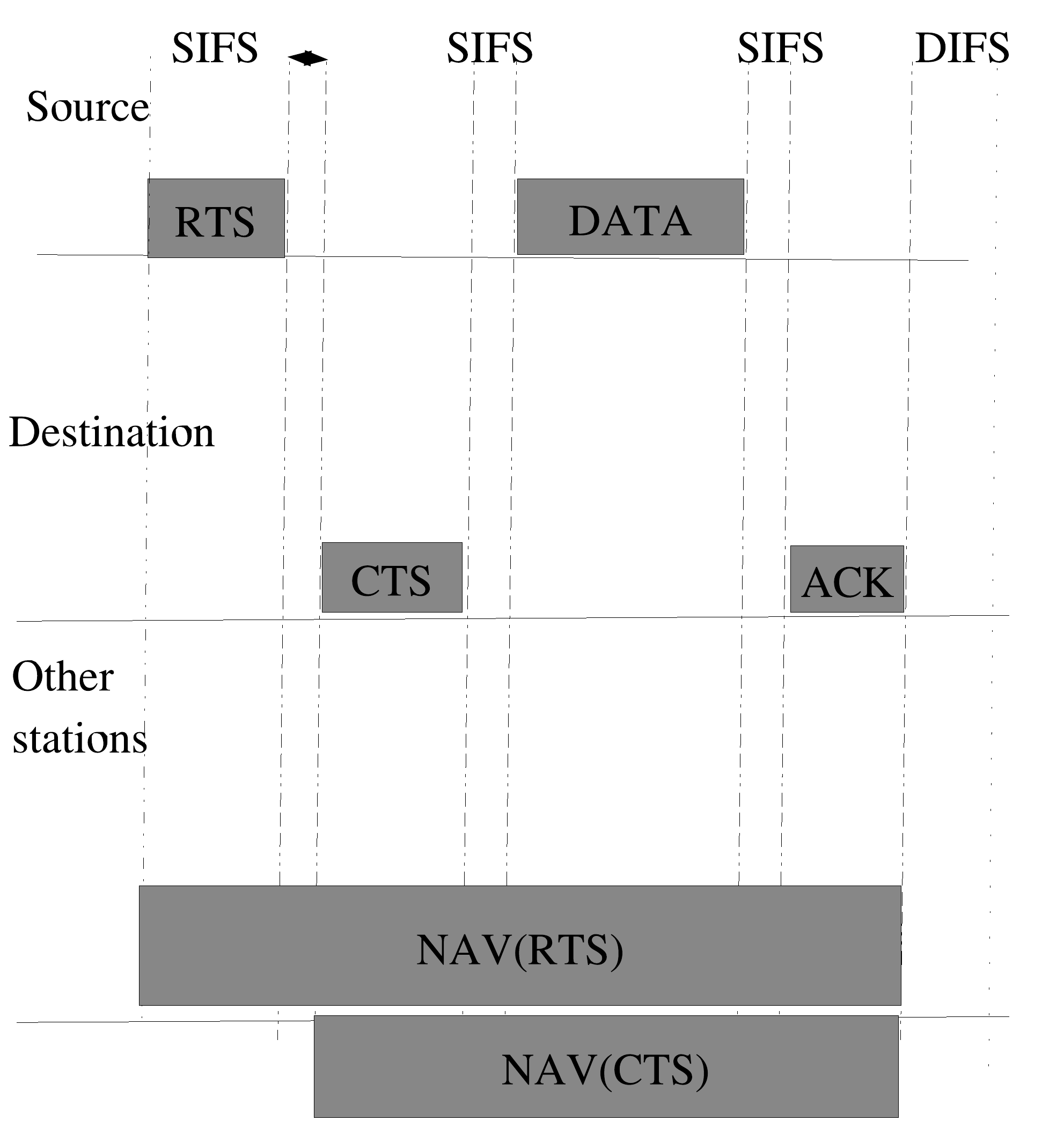}
\caption{RTS/CTS mechanism}
\label{acsqos2}

\end{figure}

%\begin{subfigure}{.5\textwidth}
%  \centering
%  \includegraphics[width=.4\linewidth]{image1}
%  \caption{A subfigure}
%  \label{fig:sub1}
%\end{subfigure}%
%
%\begin{figure}[!thb]
%\centering
%\begin{center}
%\leavevmode
%\includegraphics[width=3.5in,height=3.5in,angle=0]{rtscts.eps}
%\caption{RTS/CTS mechanism}
%\label{acsqos}
%\end{center}
%\end{figure}

The IEEE 802.11 DCF operates in two modes. First mechanism is the basic access mechanism and the second mechanism is by using RTS/CTS frames. Figure \ref{acsqos1} illustrates the basic access mechanism. In basic access mechanism a station with a new frame to transmit, continuously monitors the channel activity.  If the channel is sensed to be idle for a period equal to at least  Distributed Inter Frame  Space (DIFS) time interval the station transmits. On the other hand if the channel is sensed busy, the station continuously monitors the channel activity, until the channel becomes idle for at least DIFS time. At this point the station generates a random number of slots, the random number being uniformly distributed between (0,CW-1), where "CW" denotes the contention window size. 

The initial value of CW is set as W$_0$, a value  predefined by the standard. A back-off timer keeps track of the number of slots to wait for frame transmission. The back-off count value  is decremented by "1" in  every slot as long as the channel is idle. The counter is frozen whenever the  channel is sensed busy. The countdown is reactivated when channel is again sensed idle for more than DIFS time interval.  When the back-off timer count reaches zero the station attempts a transmission. If the transmitting station starts receiving ACK (acknowledgement) for a transmitted frame within SIFS time interval (defined by standards), the transmitting station assumes it as a successful frame transmission. Lack of an acknowledgement within SIFS time interval for a transmitted frame is assumed to indicate  an unsuccessful frame transmission.  Since  SIFS time interval is less than DIFS time interval, ACK do not encounter collision. For each unsuccessful frame transmission CW is doubled until it becomes a predefined maximum value CW$_{max}$. There after CW remains constant at CW$_{max}$. According to the standards \cite{IEEEhowto:kopka},  a frame can have a finite number of retrials with CW$_{max}$.  If collision occurs after finite retrials, the frame is dropped.

RTS/CTS mechanism is a four way hand shaking mechanism.  Figure \ref{acsqos2} illustrates the RTS/CTS mechanism wherein the source station sends a short frame called RTS before data transmission. Then the destination station responds with CTS at the end of  the RTS frame after a period of time equal to SIFS, if RTS is received correctly. On the reception of CTS the sender sends the data frame to the destination within a time duration of SIFS.    The successful reception of data frame is followed by an ACK frame after a period of time equal to SIFS. Successful exchange of RTS/CTS ensures that channel has been reserved for transmission for particular pair of sender and receiver.

The  RTS and CTS frames  carry the length of the data frame to be transmitted. All other stations hearing RTS or CTS frame update their  Network Allocation Vector (NAV) containing the information of the period of time in which the channel will remain busy. Therefore other stations delay their transmission and thereby avoid interference.  RTS/CTS mechanism is useful only when longer data frames are transmitted since for smaller data frame  the overhead caused by using RTS/CTS would degrade the performance.
%\subsection{Overview of IEEE 802.11e MAC Protocol}
 
 The IEEE 802.11e standard defines a new coordination function called the Hybrid Coordination Function (HCF) to support QoS. The contention based channel access mechanism of HCF is called as the Enhanced Distributed Channel Access  (EDCA). The centrally  controlled, contention free channel access mechanism of HCF is called the HCF controlled channel access (HCCA). The HCF combines the features of both EDCA and HCCA. We concentrate only on EDCA because HCCA is not widely used. The EDCA  supports QoS by assigning different priorities for the incoming traffic.

\begin{figure}[!thb]
\centering
\begin{center}
\leavevmode
\includegraphics[width=3.5in,height=2.25in,angle=0,trim=0mm 0mm 0mm 0mm, clip]{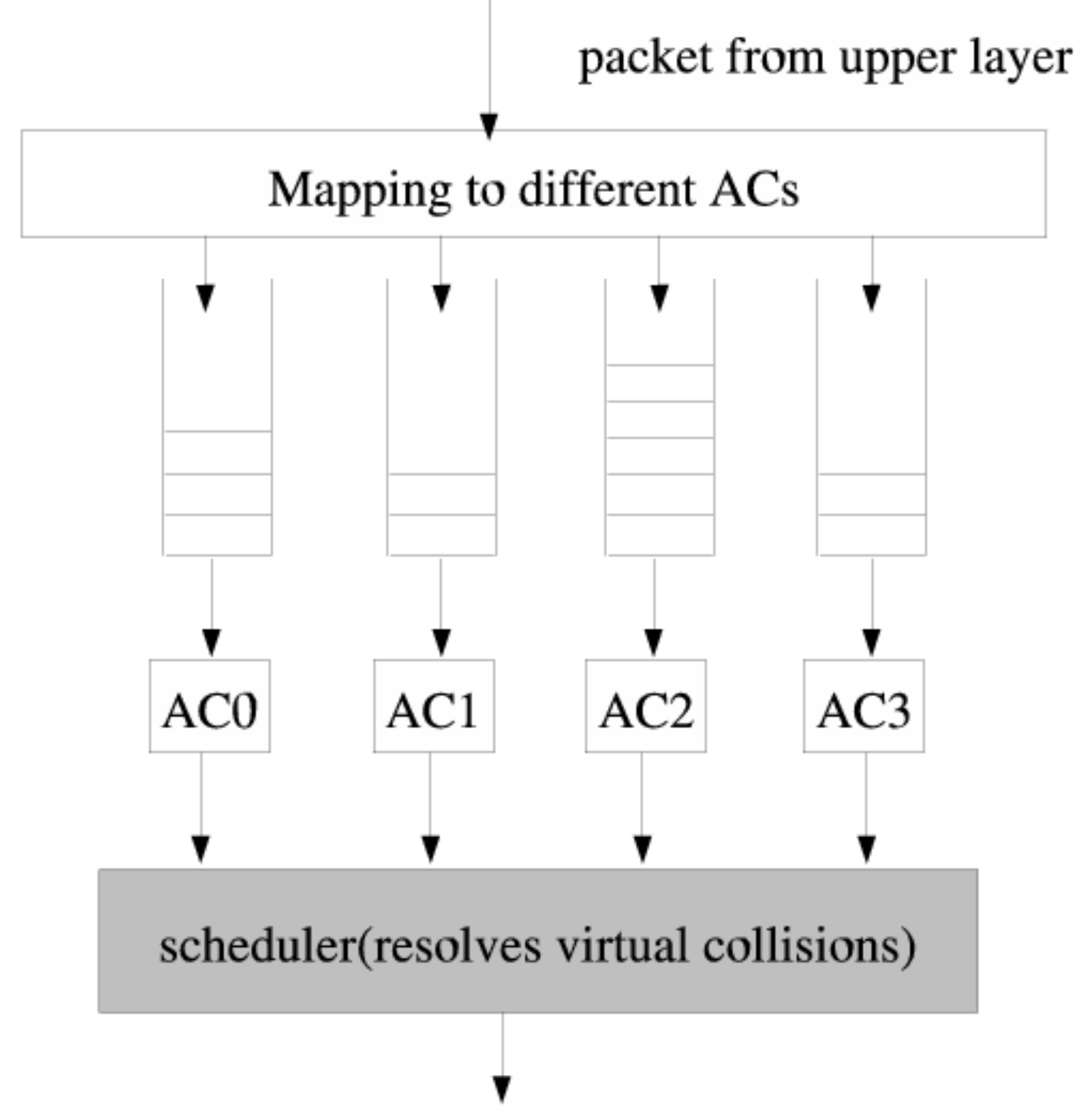}
\caption{Four different ACs of an IEEE 802.11e EDCA station}
\label{acsqos}
\end{center}
\end{figure}

Fig.\ref{acsqos} shows EDCA mechanism. As shown in Fig.\ref{acsqos},  each EDCA station maintains four queues for the four different Access Categories (ACs). Each frame with a particular kind of traffic is mapped to a particular  AC. The service differentiation is realised by using  different set of parameters for each AC.  Parameters used for service differentiation are,\newline

\begin{enumerate}

\item AIFS (Arbitration Inter Frame Spacing): This is the time period  for which medium has to be sensed idle before transmitting or initiation of backoff.

\item CW$_{max}$ and CW$_{min}$ :  These are the maximum and minimum values of contention window sizes.

\item TXOP (Transmission opportunity): The maximum duration of the frame transmission after channel is accessed.  

\end{enumerate}

\begin{table}[ht]
\centering
\begin{tabular}{|c|c|c|c|c|r|}
        \hline
$ $ & $AC$    &   $CWmin$  & $CWmax$   & AIFSN   & TXOP \\
        \hline 0   & AC0 & 7  & 15 & 2 & 3.264ms   \\
       \hline
1   & AC1 & 15 & 31 & 2 & 6.016ms\\
        \hline
2   & AC2 &31  & 1023 & 3 & 0 \\
       \hline
3   & AC3 &31  &1023  & 7  & 0\\
        \hline
\end{tabular}
\label{table11}
\caption{Parameters used for service differentiation \\ (IEEE 802.11e standard \cite{IEEEhowto:kopka})}
\end{table}

As shown in the table I, the high priority AC (i.e AC$_0$) has smaller  values of AIFS, CW$_{min}$, CW$_{max}$, and larger value of TXOP compared to low priority AC (i.e AC$_3$). There are two types of collisions that are present in EDCA. They are internal collisions (also called as virtual collisions) and external collisions. Virtual collision occurs because two or more ACs  within a station attempt to transmit in the same slot ( i.e , decrease their back-off counter to zero simultaneously). When virtual collision occurs transmission opportunity is given to the highest priority AC and the lower ACs behave as if an external collision has occurred and follows the back-off rules.  The scheduler resolves the internal collisions by assigning transmission opportunity to the highest AC involved in collision.  

\subsection{Literature Survey}
\label{litserv}
Many analytical models for DCF have been proposed in the literature \cite{IEEEhowto:bia}, \cite{ IEEEhowto:ant},  \cite{IEEEhowto:boli},   \cite{IEEEhowto:taka}, \cite{IEEEhowto:omesh},   \cite{IEEEhowto:rapti},  \cite{newmultirate}, %\cite{newpriorityscheme1}, %\cite{backoff}  
%\cite{newBEB},
%\cite{IEEEhowto:Ting}, %\cite{newsinglehop},
%\cite{IEEEhowto:finiteretrydcf}
%\cite{IEEEhowto:MG1},
%\cite{IEEEhowto:admission80211},
%\cite{xiali}, %\cite{newnitingupta},
which are relevant to our work. Bianchi \cite{IEEEhowto:bia}, proposed a two dimensional markov chain  model and  calculated the saturation throughput for IEEE 802.11 DCF. Assumptions made by Bianchi were (i) constant and independent collision probability for each transmitting frame, and  (ii) ideal channel conditions. Most of the authors in the literature use the markov chain similar to the Bianchi's markov chain.  

Xiao \cite{newpriorityscheme2} extended the Bianchi's Markov Chain model to priority schemes and compared these priority schemes with DCF. But the concept of virtual collision was not considered. Tao \cite{newpanwar} proposed a three dimensional Markov Chain for IEEE 802.11e EDCA mechanism and calculated the throughput analytically for each access categories of EDCA. Zheng \cite{newkong} proposed a Markov Chain for DCF under saturated traffic conditions and calculated the throughput by considering the concept of virtual collision and different AIFS parameter values. Yin \cite{newfoh80211e} analysed protocol service time of EDCA under statistical traffic using Markov Chain model. Throughput and average frame transmission delay have been calculated under statistical traffic conditions.

Liu \cite{newjingliu} calculated the performance of EDCA under unsaturated traffic conditions. Throughput and average frame transmission delay are calculated analytically using a four state Markov Chain.   Tinnirello et al. have modeled 802.11e EDCA that relies on decoupling approximation and performs fixed point computation of the backoff counter distribution after a generic transmission attempt \cite{Tinnirello}.
  MacKenzie et al. have modelled 802.11e as p-persistent CSMA where the access probability  can be varied for different access categories using the differentiation of CW, AIFS etc. \cite{MacKenzie}.  Qinglin et al. have considered an unsaturated 802.11e network and models joint differentiation of all four EDCA parameters for arbitrary buffer size, using M/G/1/K queuing model \cite{Qinglin}. 
  
Abu et al. have studied the performance of 802.11e EDCA under finite load and error prone channel, using a multidimensional Markov chain  along with another Markov model for channel \cite{Abu}. 
%Dapeng et al. have modeled 802.11e WLAN using a three dimensional Markov Chain wherein apart from the backoff stage and backoff counter a third dimension indicating the time remaining during either the backoff freezing period or transmission period is included \cite{Dapeng}.  
Ramaiyan et al. have performed a fixed point analysis of 802.11e WLAN and shown the presence of multistability when multiple unbalanced fixed point exist \cite{Ramaiyan}.  Inan et al. considered the analysis of the 802.11e EDCA function using a DTMC model \cite{Inan}.

In most of the papers mentioned above throughput and average frame transmission delay were calculated for different access categories without considering the concept of internal collisions.  Some models do not consider the effect of TXOP on the QoS-differentiation. Further there is a need for analyzing jitter which is an important QoS metric.  Our model differs from existing models  in the following ways :
\begin{enumerate}
%\begin{doublespace}

   \item As defined by the standards we consider the virtual collisions among different ACs inside each EDCA station in addition to external collisions.
    \item In our model we consider the effect of all the  priority  parameters such as  CW$_{max}$,  CW$_{min}$,  AIFS, TXOP and finite number of retrials.   
    \item Average frame transmission delay as well as jitter of all  ACs are derived analytically.

%\end{doublespace}      
  \end{enumerate}

This paper is organized as follows.
% Section \ref{litserv}, describes literature survey on performance analysis of IEEE 802.11 DCF and 802.11e EDCA protocol.
 Section \ref{markovmodel} describes our  modified Bianchi's Markov chain model of a DCF station valid for any one category. We derive an analytical formula for probability of transmission of a frame in any generic time slot. Then we extend our analytical results which we get in the section \ref{markovmodel} to EDCA mechanism in section \ref{edcaanalysis}. In the section \ref{delayanalysis}, we derive an analytical formula for the average frame transmission delay. In the section \ref{jitteralysis} we analyse jitter for different ACs of an EDCA station.   In the section \ref{result}, we analyse our results based on numerical computation. Finally, our conclusions are given in the section \ref{concl}.

%In this paper we propose a more general analytical model for IEEE 802.11 EDCA with virtual collisions, and we analyze effect of QoS parameters on average frame transmission delay and jitter for different ACs of an EDCA station. In the Section \ref{markovmodel}, we discuss the Markov chain model for an AC in an EDCA station. In the section \ref{analysis}, we analyze the probability of transmission of a frame by a station, considering the concept of virtual collisions. In the section \ref{delayanalysis}, we derive an analytical expression for the average frame transmission delay for an access category of an EDCA station. In the section \ref{jitteralysis} we derive an expression for jitter for different ACs for an EDCA station. In the section \ref{result}, we validate our analytical formulae for delay and jitter with NS-2 \cite{ns2} simulations. Finally our conclusions are given in the section \ref{conclusions}.

\section{Modified Bianchi's Markov Chain Model for  DCF }
\label{markovmodel}

\begin{figure}[!thb]
\centering
\begin{center}
\leavevmode
\includegraphics[width=4.4in,height=5in,angle=0,trim=50mm 5mm 0mm 5mm, clip]{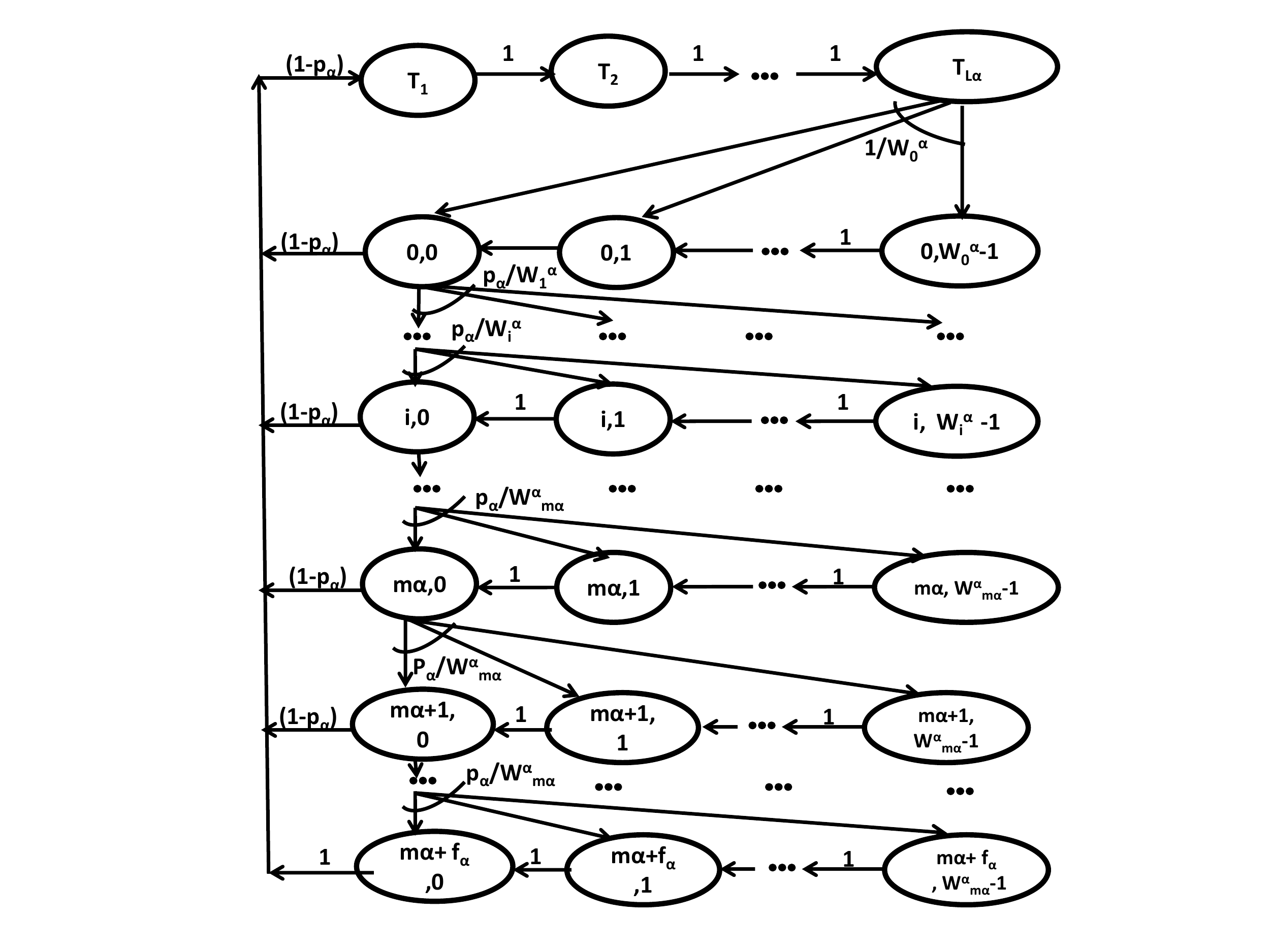}
\caption{Markov Chain model for different ACs of  EDCA station under saturated traffic conditions}
\label{figdcfsat}
\end{center}
\end{figure}

In this section we introduce and analyze a modified Bianchi's markov chain model shown in the fig.~\ref{figdcfsat}. Assumptions made are (i) constant and independent collision probability for each transmitted frame, and  (ii) ideal channel conditions. We use the index $\alpha$ to denote the access category (AC), where $\alpha$ = 0,1,2,3.  We assume that each of the ACs can be modeled as Markov chain shown in the Fig. \ref{figdcfsat}.

Let  $W_{0}^{\alpha}$ be the $CW_{min}$ of the AC number $\alpha$. Let $m\alpha$ \footnote{We have used the symbol $m\alpha$ and not $m_{\alpha}$ to denote the maximum number of stages in $AC_{\alpha}$, since $m\alpha$ itself appear as subscript in certain places.} denote the number of stages for the $AC_{\alpha}$. Let $f_{\alpha}$ denote the number of retrials permitted for $AC_{\alpha}$ after $m\alpha$ stages. Let $L_{\alpha}$ denote the number of data packets that can be transmitted with in the TXOP period by the $AC_{\alpha}$.

%As we have discussed int he section \ref{intro}, the DCF mechanism uses CSMA/CA (carrier sense multiple access with collision avoidance) mechanism with slotted Binary Exponential Back-off algorithm.  
 Following is the explanation of the  Markov Chain model shown in the Fig.~\ref{figdcfsat}. In the Fig.~\ref{figdcfsat}, the  state $\left\{i,k\right\}$ at any given slot corresponds to $i^{th}$ back-off stage $s(t)$ and with a value of back-off counter $b(t)$ as $k$. The values of $CW_{min}$ for $AC_{\alpha}$ denoted as $W_{0}^{\alpha}$ are given in table I.  The size of the collision window for the $i^{th}$ stage (reached after $i$ collisions) belonging to $AC_{\alpha}$ denoted as $W_{i}^{\alpha}$ is given by $W_{i}^{\alpha}=2^{i}W_{0}^{\alpha}$.  The variable $i$ ranges from $0$ to $m\alpha+f_{\alpha}$, where $m\alpha+f_{\alpha}$  is the maximum  retransmission limit after which if collision occurs the frame is dropped.  Here  $m\alpha$ is the maximum number of stages up to which back-off counter doubles, $m\alpha$ = $log_{2}[\frac{CW_{max}}{CW_{min}}]$  , $f_{\alpha}$ is the maximum number of retrials after stage $m\alpha$.  The index $k$ ranges from $0$ to $W_{i}^{\alpha}-1$. 
 %where $W_{i}^{\alpha}$ is the contention window size in the stage $i$ of $AC_{\alpha}$. 
 The station enters the stage $i+1$ from the stage $i$ when the transmitted frame encounters a collision.

The bidimensional process $\left\{s(t),b(t)\right\} $ is Markovian \cite{IEEEhowto:bia}.
%This is because if we consider b(t) alone the information regarding the number of collisions the frame has already experienced is not known. If we consider s(t) alone the information regarding time for the packet transmission is unknown.  
We adopt the following discrete time scale. Let $t$ and $t+1$ corresponds to the beginning of two consecutive time slots. The back-off counter is decremented at the beginning of each idle slot. This implies that  between two state transitions there can either be no packet transmission or one successful packet transmission or collision. The back-off counter decrementation is stopped when the channel is busy. So, the time interval between two successive counter decrementation varies depending on the channel condition.  

 We use the following short hand notation for one step state transition probability: 
\begin{eqnarray}
&P_\alpha\{i_1, k_1 | i_0,k_0\} \nonumber\\
&= P_\alpha \{s(t+1)=i_1,b(t+1)= k_1 | s(t)=i_0, b(t)=k_0\} \nonumber
\end{eqnarray}

 Let $b_{i,k}^{\alpha}=\lim_{t\to\infty}P_{\alpha}\left\{s(t)=i,b(t)=k \right\}$ be the stationary state occupancy  probability of Markov Chain for the $AC\alpha$.
A station transmits a frame when back-off counter value is equal to zero.  If the transmission is successful the node continues to transmit its packet from the $AC_{\alpha}$ for a duration equal to  TXOP corresponding to $AC_{\alpha}$. This is indicated by a sequence of states called transmitting states denoted as T1,T2,... $TL_{\alpha}$. The values of TXOP for different ACs are shown in table I.  Let $L_{\alpha}$ denote the number of packets that can be transmitted by $AC_{\alpha}$ in its TXOP. $L_{\alpha}$ = $\lfloor{{\frac{TXOP duration of AC_{\alpha}}{Packet length of  AC_{\alpha} }}}\rfloor$, where $\lfloor{.}\rfloor$ is the floor function. For AC2 and AC3 the standard indicates the value of TXOP as '0' (refer table I). This implies that only one packet transmission is permissible in its TXOP. In other words $L_{2}=L_{3}=1$. 

 Let "$P_{\alpha}$" be the constant conditional collision probability which is derived later in the Eqn. \ref{paci}. The non-null one step state transition probabilities for the above Markov Chain can be derived along the lines similar to that of Bianchi \cite{IEEEhowto:bia},
\begin{enumerate}
%\begin{doublespace}
\item  After successful transmission of frames in its TXOP the station enters into the state $(0,k)$ where $k\in(0,W_{0}^{\alpha}-1)$
\begin{eqnarray}
P_{\alpha}(T_{1}|i,0)&=&{(1-P_{\alpha})};\; \; i\in(0,m{\alpha}+f_{\alpha}-1), \; \nonumber \\
                       & &              \;\;\;\;\;\;\;\;\;\;\;\;\;\;\;   k\in(0,W_{0}^{\alpha}-1) \nonumber \\
          &=&\mbox{\fontsize{10}{4}\selectfont $1; \;i=m{\alpha}+f_{\alpha},k\in(0,W_{0}^{\alpha}-1)$}\nonumber
\end{eqnarray}

In the above equation, for $m{\alpha}+f_{\alpha}$, $P_{\alpha}(T_{1}|i,0)=1 $. This is because in saturated traffic conditions,  station  always has frames ready for transmission. In the state, $m{\alpha}+f_{\alpha}$ there are only two possible conditions. They are successful frame transmission after which station enters into the stage "$0$", with probability "$1-P_{\alpha}$" or dropping of the frame with probability "$P_{\alpha}$". In either case station enters into the stage "$0$".

\item During the TXOP period packets are transmitted one after another  until $L_{\alpha}$ packets get transmitted.
\begin{eqnarray}
P(T_{i+1}|T_{i})=1;\; \;  i\in(0,L_{\alpha}-1),  \; \;   
                                  for \; \;  \alpha=0,1 
\end{eqnarray}
Followed by this, the backoff process is initiated by setting the counter to a random value following the distribution $U(0,W_{0}-1)$. Recall that for $AC_{2}$ and $AC_{3}$ $L_{\alpha}=1$.
\begin{eqnarray}
P(0,k|T_{L_{\alpha}})&=&\frac{1}{W_{0}^{\alpha}}\; \; ; \; \;   \alpha= 0,1\; \nonumber
\end{eqnarray}
\begin{eqnarray}
P(0,k|T_{1})&=&\frac{1}{W_{0}^{\alpha}}\; \; ; \; \; \alpha= 2,3\; \nonumber
\end{eqnarray}
\item Stage changes from $i-1$ to $i$ when a transmitted frame encounters collision.
\begin{eqnarray}
P_{\alpha}(i,k|i-1,0)&=&\frac{P_{\alpha}}{W_{i}^{\alpha}};\;\;i\in(1,m\alpha),\; \nonumber \\
                       & &              \;\;\;\;\;\;\;\;\;\;\; k\in(0,W_{i}^{\alpha}-1)  \nonumber\\
            &=&\frac{P_{\alpha}}{W_{i}^{\alpha}};\;\;i\in(m{\alpha} +1,m\alpha+f_{\alpha}),\nonumber\\
		& &\;\;\;\;\;\;\;\;\;\;\;k\in(0,W_{m\alpha}^{\alpha}-1) \nonumber
\end{eqnarray}
\item Back-off time counter is decremented by "1" in each slot. 
\begin{eqnarray}
P_{\alpha}(i,k|i,k+1)&=1;\;\;i\in(1,m\alpha),\;\; 
                         k\in(0,W_{i}^{\alpha}-2)  \nonumber
\end{eqnarray}
%\end{doublespace}
\end{enumerate}
The effect of freezing the back-off counter (for busy channel conditions) is considered later in the section III.

 In the steady state the following equations can be derived along the lines similar to that of Bianchi \cite{IEEEhowto:bia}: \newline
 The expression for $b_{i,0}^{\alpha}$ is given by,
\begin{eqnarray}
\label{b00}
b_{i,0}^{\alpha}=b_{0,0}^{\alpha}P_{\alpha}^i, \;\;\;\;\;i\in(0,m{\alpha}+f_{\alpha})
\end{eqnarray}
From the inspection of Markov chain, the expression for $b_{i,k}^{\alpha}$ can be written as,
\begin{eqnarray}
b_{i,k}^{\alpha}=\frac{(W_{i}^{\alpha}-k)}{W_i^{\alpha}} b_{i,0}^{\alpha}  \;\;  i\in(0,m{\alpha}-1) \;\;  k\in (0,W_{i}^{\alpha}-1) 
\label{bik1}
\end{eqnarray}
Now we derive an expression for $b_{0,0}^{\alpha}$ in terms of "$P_{\alpha}$".
From the probability conservation formula\cite{ross2014introduction},
\begin{eqnarray}
\label{sumprob}
\sum_{j=0}^{L_{\alpha}}T_{j}+\sum_{i=0}^{m{\alpha}+{f_{\alpha}}}\sum_{k=0}^{W_{i}^{\alpha}-1}b_{i,k}^{\alpha} =1.
\end{eqnarray}
substituting Eqns. \ref{b00}, \ref{bik1} in Eqn.\ref{sumprob} and noting that $P(T_{j})$ = $b_{0,0}$ and $j$ = $1$ to $L_{\alpha}$ we get,
\begin{eqnarray}
\label{boo123}
b_{0,0}^{\alpha}\Bigg[{\frac{L_{\alpha}}{W_{0}^\alpha}}+\sum_{i=0}^{m\alpha}\left(\frac{W_{i}^{\alpha}+1}{2}\right)P_{\alpha}^{i}+  \nonumber    
\\
\sum_{i=m\alpha+1}^{m\alpha+f_\alpha}\left(\frac{W_{m\alpha}^{\alpha}+1}{2}\right)P_{\alpha}^i \Bigg] =1
\end{eqnarray}
From the Eqn.\ref{boo123}, after some algebra we get an expression for $b_{0,0}^{\alpha}$ given by the Eqn.\ref{eqn_dbl_j}.   

%  \begin{figure*}[!t]
%  % ensure that we have normalsize text
%  \normalsize
%    %\setcounter{equation}{13}
%	\begin{equation}
%	\label{eqn_dbl}
%	 b_{0,0}=\mbox{\fontsize{18}{4}\selectfont $\frac{1}{\left[\left(\frac{1-(2P)^{m+1}}{1-2P}\right)\frac{W_0}{2}  
%      + \left(\frac{1-P^{m+1}}{1-P}\right)\frac{1}{2}+P^{m+1}\left(\frac{W_m+1}{2}\right)\left(\frac{1-P^f}{1-P}\right)
%\right]}$} 
%	 \end{equation}
%     \end{figure*}

Now we derive an analytical formuale for probability of transmission of a frame in any generic time slot "$\tau_{\alpha}$". Since a frame will be transmitted when the counter value is equal to zero, the expression for "$\tau_{\alpha}$" is given by the equation,
 
\begin{eqnarray}
%\tau=\sum_{i=0}^{m+f}b_{i,0}=\sum_{i=0}^{m+f}P^ib_{0,0}=b_{0,0}\left[\frac{1-P^{m+f+1}}{1-P}\right].
\tau_{\alpha}=\sum_{i=0}^{{m\alpha}+{f_{\alpha}}}b_{i,0}^{\alpha}.
\label{tausat}
\end{eqnarray}
Where $b_{i,0}^{\alpha}$ is the state occupancy probability of a station in a generic stage "i". The expression for $b_{i,0}^{\alpha}$ is given by the Eqn. \ref{b00}.

On substituting Eqn.\ref{b00} in Eqn.\ref{tausat} we get,
\begin{equation}
\tau_{\alpha}=\sum_{i=0}^{{m\alpha}+{f_{\alpha}}}b_{i,0}^{\alpha}=\sum_{i=0}^{{m\alpha}+{f_{\alpha}}}P_{\alpha}^{i}b_{0,0}^{\alpha}=b_{0,0}^{\alpha}\left[\frac{1-P_{\alpha}^{{m\alpha}+{f_{\alpha}}+1}}{1-P_{\alpha}}\right].
%\tau=\sum_{i=0}^{m+f}b_{i,0}.
\label{tausat11}
\end{equation}

\section{Analysis of IEEE 802.11e EDCA with virtual collisions}
\label{edcaanalysis}
In this section we extend the results  which we got in the section \ref{markovmodel} to EDCA mechanism by considering the concept of virtual collisions. We derive an analytical expression for probability of transmission of a frame in any generic time slot for an EDCA station.\newline 

As shown in the Fig. \ref{acsqos}, EDCA maintains four different ACs. Each of the ACs has different QoS parameters. 
Each AC contends for the channel with different values of AIFS and different sizes of $CW_{max}$ and $CW_{min}$. Let $AIFS_\alpha$ be the value of AIFS for an AC "$\alpha$". According to the standards the value of $AIFS_\alpha=SIFS+AIFSN_\alpha*aslottime$ where "$AIFSN$" is Arbitrary Inter Frame Space Number which is different for different AC. The values of AIFSN for four ACs are given in the table I. 

 When virtual collision occurs within an EDCA  station, transmission opportunity is given to the highest colliding  AC, and lower AC differ their transmission and follow the back-off rules.  Let $\tau_0,\tau_1,\tau_2$ and $\tau_3 $ be the probabilities of transmission of a frame from AC$_0 $, AC$_1$, AC$_2$ and AC$_3$  respectively as given by Eq.\ref{tausat11} for $\alpha=0,1,2,3$. The probability of transmission of a frame by an EDCA station from any AC in any generic time slot "$\tau$" is given by Eqn.\ref{tauoverall}.

% From the Eqn. \ref{tausat11}, the expression for $\tau_j$ for an $AC_j$ is given by,
%
%\begin{eqnarray}
%\label{tauj}
%\tau_j=\sum_{i=0}^{m+f}b_{i,0,j}=\sum_{i=0}^{m+f}(P_j)^ib_{0,0,j}=b_{0,0,j}\left[\frac{1-(P_j)^{m+f+1}}{1-P_j}\right]
%\end{eqnarray}
%
%The expression for $b_{0,0,j}$  is given by \ref{eqn_dbl_j},

 \begin{figure*}[!t]
  % ensure that we have normalsize text
  \normalsize
	\begin{equation}
	\label{eqn_dbl_j}
b_{0,0}^{\alpha} = \mbox{\fontsize{18}{4}\selectfont $\frac{1}{\left[{\frac{L_{\alpha}}{W_{0}^\alpha}}+\left(\frac{1-(2P_{\alpha})^{({m{\alpha}+1)}}}{1-2P_{\alpha}}\right)\frac{W_{0}^{\alpha}}{2}  
      + \left(\frac{1-(P_{\alpha})^{({m{\alpha}+1})}}{1-P_{\alpha}}\right)\frac{1}{2}+(P_{\alpha})^{({m{\alpha}+1})}\left(\frac{W_{m\alpha}^{\alpha}+1}{2}\right)\left(\frac{1-(P_{\alpha})^{f_{\alpha}}}{1-P_{\alpha}}\right)\right]}$} 
	 \end{equation}
     \end{figure*}

\begin{figure*}[!t]
\normalsize
\begin{equation}
\label{tauoverall}
 \mbox{\fontsize{12}{4}\selectfont $\tau=\tau_0+\tau_1(1-\tau_0)+\tau_2(1-\tau_0)(1-\tau_1)+\tau_3(1-\tau_0)(1-\tau_1)(1-\tau_2)$}.
\end{equation}
\end{figure*}

Now let us derive an expression for probability of collision of a frame from an AC$_{\alpha}$. Let $PI_{\alpha}$ represents the probability of internal collision for a frame from an AC$_{\alpha}$.
\begin{enumerate}
%\begin{doublespace}
\item Internal collision will not occur for a frame which has been transmitted from AC$_0$.
\begin{eqnarray}
\label{pi0}
PI_0=0 
\end{eqnarray}
\item Internal collision for a frame  in AC$_1$ will happen only when a frame from AC$_0$ transmits in the same slot.
\begin{eqnarray}
\label{pi1}
PI_1=\tau_0
\end{eqnarray}
\item Internal collision for AC$_2$ will happen when a frame from any one of the AC$_0$ or AC$_1$ transmits in the same slot.
\begin{eqnarray}
\label{pi2}
PI_2=1-(1-\tau_0)(1-\tau_1)
\end{eqnarray}
\item Internal collision for AC$_3$ will happen when a frame from any one of the  AC$_0$ or AC$_1$ or AC$_2$  transmits in the same slot.
\begin{eqnarray}
\label{pi3}
PI_3=1-(1-\tau_0)(1-\tau_1)(1-\tau_2)
\end{eqnarray}
%\end{doublespace}
\end{enumerate}
The probability of collision for AC$_{\alpha}$, $P_{\alpha}$  is given by,
\begin{eqnarray}
 \label{paci}
 P_{\alpha}=PI_{\alpha}+(1-PI_{\alpha})P_{ex}.
\end{eqnarray}
Where P$_{ex}$ is the probability that the transmitted frame encounters an external collision which is same for all stations. The expression for  P$_{ex}$ is given by the equation,
\begin{eqnarray} 
\label{pex}
P_{ex}=1-(1-\tau)^{n-1}.
\end{eqnarray}
%where "n" is the number of stations.

We now derive  P$_{S\alpha}$  the conditional probability that the frame transmitted from the AC$_\alpha$ is successful given that a frame is transmitted by a station.

\begin{enumerate}
%\begin{doublespace}
\item Let P$_{S0}$ be the conditional probability that the frame transmitted from AC$_0$ is successful given that a frame is transmitted by a station. This is possible when no other station transmits in the same slot. The expression for P$_{S0}$ is given by,
\begin{eqnarray}
\label{ps0}
P_{S0}=\frac{n\tau_0(1-\tau)^{n-1}}{1-(1-\tau)^n}
\end{eqnarray}
\item Let P$_{S1}$ is the conditional probability that a frame transmitted from AC$_1$ is successful given that a frame is transmitted by the station. This is possible when (i) a packet from AC$_0$ of the same station is not transmitted in the same slot and (ii) no other station transmits in the same slot. The expression for P$_{S1}$ is given by, 
\begin{eqnarray}
\label{ps1}
P_{S1}=\frac{n\tau_1(1-\tau_0)(1-\tau)^{n-1}}{1-(1-\tau)^n}
\end{eqnarray}
\item Let P$_{S2}$ be the conditional probability that the transmission of frame from AC$_2$ is successful. This is possible when (i) a frame from either AC$_0$, AC$_1$ belonging the same station is not transmitted in the same slot and (ii) no other station transmits in the same slot. The expression for P$_{S2}$ is given by,  
\begin{eqnarray}
\label{ps2}
P_{S2}=\frac{n\tau_2(1-\tau_0)(1-\tau_1)(1-\tau)^{n-1}}{ 1-(1-\tau)^n}
\end{eqnarray}
\item Let P$_{S3}$ be the conditional probability that the transmission of the frame from AC$_3$ is successful. This is possible when (i) a frame from either AC$_0$, AC$_1$, AC$_2$ belonging the same station is not transmitted in the same slot and (ii) no other station is transmitted in the same slot. The expression for P$_{S3}$ is given by, 
\begin{eqnarray}
\label{ps3}
P_{S3}=\frac{n\tau_3(1-\tau_0)(1-\tau_1)(1-\tau_2)(1-\tau)^{n-1}}{1-(1-\tau)^n}.
\end{eqnarray}
%\end{doublespace}
\end{enumerate}
From the simultaneous non-linear  Eqns.  from \ref{tausat11} to \ref{pex} by using numerical methods  we can find the value of probability of transmission of a frame in any generic slot  "$\tau$" for a given EDCA station.

%The Markov Chain shown in the fig.\ref{figdcfsat} for each of the access category  of an EDCA station is extended from the Bianchi's  two dimensional Markov Chain model . We accommodate the QoS features of 802.11e EDCA  for each of the ACs. 
\section{Average frame transmission delay analysis}
\label{delayanalysis}

In this section, we derive an analytical expression for average frame transmission delay for different ACs of an EDCA station. \newline 

Let "$N^{\alpha}$" be the number of states visited by a station for successful packet transmission from an $AC_{\alpha}$ (i.e, number of  back-off counter  decrements for successful packet transmission). The expected number of slots visited by  a station belonging  an AC$_{\alpha}$, in the stage $i$ is $E(N_{i}^{\alpha})+1 $, where  $N_{i}^{\alpha}$ is $ U(0,W_{i}^{\alpha}-1) \;\;\;\;$ then $\;\;\;\; E(N_{i}^{\alpha})=\displaystyle{\frac{W_{i}^{\alpha}-1}{2}}$. One extra slot accounts for the frame transmission.
\begin{equation}
\label{enacixx}
E(N^{\alpha})=L_{\alpha}+\sum^{m{\alpha}+f_{\alpha}}_{i=0}\left(E(N_{i}^{\alpha})+1\right)\frac{(P_{\alpha})^i(1-P_{\alpha})}{(1-(P_{\alpha})^{m{\alpha}+f_{\alpha}+1})} 
\end{equation}
where $\displaystyle{\frac{(P_{\alpha})^i(1-P_{\alpha})}{(1-(P_{\alpha})^{m{\alpha}+f_{\alpha}+1})}}$ is the probability that transmission is successful in the stage $i$ for an $AC_{\alpha}$,  given that the frame is not dropped. The term $L_{\alpha}$ in equation \ref{enacixx} accounts for the number of slots spent in the TXOP period. $ E(N^{\alpha}) $ is different for different ACs because,  $ E(N^{\alpha}) $  is a function of minimum and maximum values of contention window,  $W_{0}^{\alpha}$ and $W_{m{\alpha}}^{\alpha}$ which is different for different ACs.

\subsection{Derivation of expected time duration of the state transition for AC$_{\alpha}$:}

Let T$_{s}^{\alpha}$  be the time the channel is sensed busy because of  successful frame transmission, and  T$_{c}^{\alpha}$ be the time  the channel is sensed busy by each station because of collision for a frame from an AC$_{\alpha}$.  For the basic access mechanism, value of T$_{s}^{\alpha}$ for an AC$_{\alpha}$ is given by the equation, 
\begin{eqnarray}
\label{tsibasic}
 T_{s}^{\alpha} =AIFS_{\alpha}+t_H+t_{L_P}+SIFS+t_{ACK}
\end{eqnarray}
Where  t$_{L_p}$ is the time required for frame transmission which is calculated as $\displaystyle{\frac{L_p}{R}}$ where "R" is data rate and L$_p$ is the payload size . t$_H$ the time required for transmission of both PHY and MAC headers which is given by $\displaystyle{\frac{H}{R}}$ where "H" is equal to sum of MAC header and PHY header in bits. In the above equation t$_{ACK}$ is equal to $\displaystyle{\frac{L_{ACK}}{R}}$ where L$_{ACK}$ is length of ACK frame.\newline

The value of T$_{c}^{\alpha}$ for basic access mechanism is given by the equation,
\begin{eqnarray}
\label{tcibasic}
 T_{c}^{\alpha}=AIFS_{\alpha}+t_H+t_{L_P}
\end{eqnarray}
The value of T$_{s}^{\alpha}$ for RTS/CTS mechanism is given by,
\begin{eqnarray}
\label{tsrtsi}
T_{s}^{\alpha}&=&AIFS_{\alpha}+t_{RTS}+SIFS+t_{CTS}+SIFS \nonumber \\
&+&t_H+t_{L_P}+SIFS+t_{ACK}
\end{eqnarray}
and the value of  T$_{c}^{{\alpha}}$ for RTS/CTS is given by the  equation,
\begin{eqnarray}
\label{tcrtsi}
T_{c}^{\alpha}=AIFS_{\alpha}+t_{RTS}. 
\end{eqnarray}

Let E(T$^{\alpha}$) be the expected time duration of the state transition for AC$_{\alpha}$.  The channel may be idle or busy. The duration of the state transition is $T_{s}^{\alpha}$ with the probability $P_{s\alpha}$ if the channel is busy because of successful frame transmission. The value of $P_{s\alpha}$ for $\alpha=0,1,2,3$ and given by equations \ref{ps0} to \ref{ps3}.   The duration is equal to $T_{c}^{\alpha}$ with probability $P_{c{\alpha}}$ if the channel is busy because of collision for corresponding AC "${\alpha}$". If the channel is idle then station takes $slottime$ to decrement its back-off count. The expression for $E(T^{\alpha})$ is given by the Eqn. \ref{eqn_d_xyz}. Here $P_{tr}$  is the probability of a transmission in a given slot and is given by $P_{tr}=1-(1-\tau)^n$. The final term $(AFIS_{\alpha}-AIFS_{min})$ indicates  how many extra  time slots the station has to wait after the channel is sensed idle in order to decrement the back-off counter value. Numerically it is equal to the number of slots in the  DIFS time interval. According to the standards $AIFS_{min} = SIFS+ 2*slottimes $ .

From the table I, AIFSN values for AC$_0$ and AC$_1$ is equal to 2,  for AC$_2$ has a value of 3 and AC$_3$ has a value of 7. So, AC$_0$ and AC$_1$ decrement the back-off count value by one if the channel is idle for DIFS time interval. AC$_3$ has to wait for an one slot extra to decrement its back-off count and AC$_3$ has to wait 5 more slot times to decrement its back-off count as compared to AC$_0$ and AC$_1$.

Note that every time  a transmission attempt become successful $L_{\alpha}$ packets will be transmitted in the  TXOP by $AC_{\alpha}$.  Thus the  channel access delay  "$D^{\alpha}$" for one packet belonging to  $AC_{\alpha}$ is given by,
\begin{eqnarray}
\label{backoffdelay}
D^{\alpha} =  \frac{ N^{\alpha} * E(T^{\alpha})}{L_{\alpha}}
\end{eqnarray}

  %The expression for E(T$^{\alpha}$) is given by the Eqn.\ref{eqn_dbl_tt}.

 \begin{figure*}[!t]
 \normalsize
 \begin{equation}
  \label{eqn_d_xyz}
E(T_{\alpha})=(1-P_{tr})\sigma+P_{tr}\left[(AIFS_{\alpha}-AIFS_{min})+ \sum^{3}_{{\alpha}=0}P_{s{\alpha}}T_{s}^{\alpha}+\sum^{3}_{\alpha = 0}(1-P_{s{\alpha}})T_{c}^{\alpha}\right]
	 \end{equation}
	  % Restore the current equation number.
          %\setcounter{equation}{\value{MYtempeqncnt}}
	 % IEEE uses as a separator
	 \hrulefill
	% The spacer can be tweaked to stop underfull vboxes.
	\vspace*{4pt}
	\end{figure*}

From the Eqn. \ref{backoffdelay}, the  expected frame transmission delay for an $AC_{\alpha}$, "$E(D^{\alpha})$" is given by,
\begin{eqnarray}
\label{edsat}
E(D^{\alpha})=\frac{E(N^{\alpha}).E(T^{\alpha})}{L_{\alpha}}
\end{eqnarray}

$E(T^{\alpha})$ is given by the Eqn. \ref{eqn_d_xyz} and $E_(N^{\alpha})$ is given by the Eqn. \ref{eqn_dbl_tt}.

\section{Jitter Analysis}
\label{jitteralysis}

The  Jitter  is defined as the square root of variance in the delay. , $J =\sqrt (Var(D))$. Let  Var(D$^{\alpha}$) represents the variance in the delay for an access category  "${\alpha}$". Let  $Var(N^{\alpha})$  denote the variance of number of states visited by a frame from an $AC_{\alpha}$ for a given station for successful frame transmission.

%Var(D$^{\alpha}$)
%\begin{eqnarray}
%Var(N_{ACn})&=&E(N_{ACn}^2)-E^2(N_{ACn}) \nonumber \\
%      &=&\mbox{\fontsize{9}{4}\selectfont $\sum^{m+f}_{i=0}E(N_i^2)\frac{(P^i)(1-P)}{(1-P^{m+f+1})}-E^2(N)$} \nonumber \\
%&=&\sum^{m+f}_{i=0} [Var(N_i)+E^2(N_i)]  \frac{(P^i)(1-P)}{(1-P^{m+f+1})}-E^2(N)   \nonumber\\
%               & =&\sum^{m+f}_{i=0} \left[\frac{W_i^2-1}{12}+\left(\frac{W_i-1}{2}\right)^2\right]\frac{(P^i)(1-P)}{(1-P^{m+f+1})}-E^2(N).  \nonumber
%  \label{varN}
%\end{eqnarray}

From equation \ref{backoffdelay} treating $E(T^{\alpha})$ and $L_{\alpha}$ to be constants we get\cite{ross2014introduction},
\begin{eqnarray}
\label{variancesati}
Var(D^{\alpha})=\frac{Var(N^{\alpha}).E^2(T^{\alpha})}{(L_{\alpha})^2}
\end{eqnarray}
$E(T^{\alpha})$  is given in the Eqn. \ref{eqn_d_xyz}. $Var(N^{\alpha})$ is given by the Eqn. \ref{varttt}.   From the Eqn.\ref{variancesati} we can find the jitter  $J_{\alpha}$ for $AC_{\alpha}$ is given by 
\begin{equation}
J_{\alpha}=\sqrt{Var(D^{\alpha})}
\end{equation}

%\section{Derivation of average frame transmition delay from an ACs}
 \begin{figure*}[!t]
  % ensure that we have normalsize text
  \normalsize
  % Store the current equation number.
 % \setcounter{MYtempeqncnt}{\value{equation}}
  % Set the equation number to one less than the one
  % desired for the first equation here.
   % The value here will have to changed if equations
   % are added or removed prior to the place these
  %quations are referenced in the main text.
 %\setcounter{equation}{17}
Expression for expected frame transmission delay of an AC of an EDCA station is given by:
	\begin{equation}
	\label{eqn_dbl_tt}
	 E(N^{\alpha})= \sum^{m{\alpha}+f_{\alpha}}_{i=0}\left(\frac{W_{i}^{\alpha}-1}{2}+1\right).\frac{(P_{\alpha})^i(1-P_{\alpha})}{1-(P_{\alpha})^{(m{\alpha}+f_{\alpha}+1)}} \nonumber 
	\end{equation}
	\begin{equation}
=\mbox{\fontsize{13}{4}\selectfont $\left[ \left(\frac{1-(2P_{\alpha})^{(m{\alpha}+1)}}{1-2P_{\alpha}}\right) \frac{W_0^{\alpha}}{2} + \\
        \frac{1}{2} \left( \frac{1-(P_{\alpha})^{(m{\alpha}+1)}}{1-P_{\alpha}}\right)+\left(\frac{W_{m{\alpha}}^{\alpha}+1}{2}\right)(P_{\alpha})^{(m{\alpha}+1)} \left(\frac{1-(P_{\alpha})^{f_{\alpha}}}{1-P_{\alpha}}\right)\right] \frac{(1-P_{\alpha})}{(1-(P_{\alpha})^{{(m{\alpha}+f_{\alpha}+1)}}}$}  
	 \end{equation}
	\end{figure*}
	
         \begin{figure*}[!t]
\hrulefill \newline
Expression for Variance of frame transmission delay:	
	\begin{equation}
	\label{eqn_dbl_y}
\mbox{\fontsize{13}{4}\selectfont $Var(N^{\alpha})=E({(N^{\alpha})}^2)-E^2(N^{\alpha})$} \nonumber
	 \end{equation}

        \begin{equation}
	\label{eqn_d_x}
=\mbox{\fontsize{13}{4}\selectfont $\sum^{m{\alpha}+f_{\alpha}}_{{\alpha}=0}{(N_{i}^{\alpha})^2}\frac{(P_{\alpha}^i)(1-P_{\alpha})}{(1-(P_{\alpha})^{(m{\alpha}+f_{\alpha}+1)})}-E^2(N^{\alpha})$} \nonumber 
	 \end{equation}
        \begin{equation}
	\label{eqn_d_x}
=\sum^{m{\alpha}+f_{\alpha}}_{i=0} [Var(N_{i}^{\alpha})+E^2(N_{i}^{\alpha})]  \frac{(P_{\alpha}^i)(1-P_{\alpha})}{(1-(P_{\alpha})^{(m{\alpha}+f_{\alpha}+1)})}-E^2(N^{\alpha})   \nonumber\\
	 \end{equation}
	\begin{equation}
	\label{varttt}
 Var(N^{\alpha})=\sum^{m{\alpha}+f_{\alpha}}_{i=0} \left[\frac{{(W_{i}^{\alpha})}^2-1}{12}+\left(\frac{({W_{i}^{\alpha})}-1}{2}\right)^2\right]\frac{(P_{\alpha}^i)(1-P_{\alpha})}{(1-(P_{\alpha})^{(m{\alpha}+f_{\alpha}+1)})}-E^2(N^{\alpha})
	\end{equation}
	%\begin{equation}
	%=\mbox{\fontsize{10}{4}\selectfont $\left[\left(\frac{(1-(4P_{ACi})^{m+1})}{1-4P_{ACi}}\right)\frac{W_m^2}{3}
	%+ \frac{W_m^2}{3}P_{ACi}^{m+1}\left(\frac{1-P_{ACi}^f}{1-P_{ACi}}\right) -
	%\left(\frac{(1-(2P_{ACi})^{m+1})}{1-2P_{ACi}}\right)\frac{W_m}{2} 
 	%+ \frac{W_m}{2}P_{ACi}^{m+1}\left(\frac{1-P_{ACi}^f}{1-P_{ACi}}\right)
   %+ \frac{1}{6}\frac{(1-P_{ACi}^{m+f+1})}{1-P_{ACi}} \frac{(1-P_{ACi})}{(1-P_{ACi}^{m+f+1})}\right] \\
%-E^2(N)$}
%	\end{equation}
	  % Restore the current equation number.
          %\setcounter{equation}{\value{MYtempeqncnt}}
	 % IEEE uses as a separator
	 \hrulefill
	% The spacer can be tweaked to stop underfull vboxes.
	\vspace*{4pt}
	\end{figure*}

\section{Numerical Result}
\label{result}
 In this section we present our analytical results computed numerically using the Mathematica package \cite{mathematica10}. 
%In this section we analyse our results for average frame transmission delay and Jitter for two different ACs of EDCA station.

%\subsection{Simulation Setup}
%\label{result1}

\begin{figure}[!h]
\centering
\begin{center}
\leavevmode
\includegraphics[width=4in,height=2.5in,angle=0]{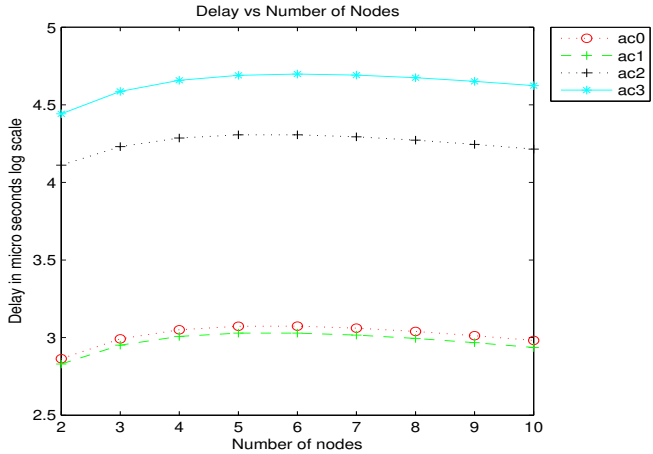}
\caption{Analytical results for delay vs number of stations for different ACs}
%('*' represents the analytical result output, 'x' represents the NS2 simulation output for AC$_3$, 'v' represents the analytical output, and '+' represents the NS2 simulation output for AC$_0$.)}
\label{80211edel}
\end{center}
\end{figure}
Fig. \ref{80211edel} shows the delay computed numerically using equations (\ref{eqn_d_xyz}), (\ref{edsat}), (\ref{eqn_dbl_tt}) for different access categories.  From Fig. \ref{80211edel} it can be seen  that the delay of access categories  $AC_{0}$ and  $AC_{1}$ are considerably less than that of $AC_{2}$ and  $AC_{3}$. Notice from Table \ref{table11}  that the values of  $CW_{min}$ and   $CW_{max}$ of  $AC_{0}$ and  $AC_{1}$  are much lower as compared to $AC_{2}$ and  $AC_{3}$ (31 - 1023). This  ensures that the expected value of the counter  $(E[b(t)])$  at any given $t$ is less for $AC_{0}$ and  $AC_{1}$  as compared to $AC_{2}$ and  $AC_{3}$. Further lower values of $AIFSN$ implies that  $AC_{0}$ and  $AC_{1}$  has to wait for lesser number of slots to restart the backoff process  once the channel becomes free. Further larger values $TXOP$ for $AC_{1}$ followed by  $AC_{0}$  allows a number of packets to be transmitted after each successful contention which reduces the contention overhead. All these factors along with the concept of internal collision contributes to the lower values of delay for $AC_{0}$ and  $AC_{1}$. The differentiating factor between $AC_{2}$ and  $AC_{3}$ are  $AIFSN$ and  the internal collision which explains the larger value of delay $AC_3$ as compared to $AC_1$.

\begin{figure}[!thb]
\centering
\begin{center}
\leavevmode
\includegraphics[width=4in,height=2.5in,angle=0]{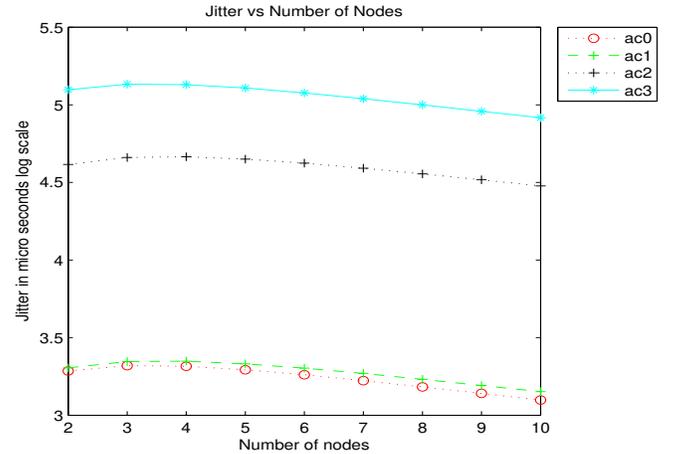}
\caption{Analytical result for Jitter vs number of stations for different ACs in IEEE 802.11e EDCA}
\label{jitvsn802.11e}
\end{center}
\end{figure}
Fig. \ref{jitvsn802.11e} shows the jitter computed numerically using equations (\ref{eqn_d_xyz}), (\ref{variancesati}),  (\ref{varttt}).
Fig. \ref{jitvsn802.11e} shows  that the jitter of access categories  $AC_0$ and  $AC_1$ are considerably less than that of $AC_2$ and  $AC_3$. The arguments mentioned above explaining lower values of delay for   $AC_0$ and  $AC_1$  as compared to $AC_2$ and  $AC_3$  also hold good for jitter. Both delay and jitter do not vary much with the number of nodes as nodes are assumed to be saturated.

\section{Conclusion}
\label{concl}
In this paper we have presented the performance of IEEE 802.11e EDCA standard under saturated and ideal channel conditions. We have derived the analytical formulae for the average frame transmission delay and jitter for different access categories of EDCA station. In our model, as defined by the standards we have considered the internal collisions which may occur within an EDCA station,  as well as external collisions which may occur among the EDCA stations. The effect of priority parameters such as CW$_{min}$, CW$_{max}$ sizes, AIFS values, and finite retrials on packet transmission delay and Jitter for an EDCA station under saturated traffic conditions have been analysed.  Our analytical results quantifies the goal of the IEEE 802.11e standard to provide QOS for high priority multimedia traffic. 
%We conclude that IEEE 802.11e standard provides traffic differentiation and QoS guarantee for multimedia traffic. 

\appendices
\ifCLASSOPTIONcaptionsoff
  \newpage
\fi

\bibliographystyle{IEEEtran}
\bibliography{references}

%\begin{IEEEbiography}{Venkata Sitaram .A.}
%\end{IEEEbiography}

%\begin{IEEEbiographynophoto}{Venkatesh . T. G.}

%\newpage

%\begin{IEEEbiographynophoto}{Manivasakan. R.}

\end{document}